\documentclass[11pt,a4paper]{article}
\usepackage{jheppub}

\usepackage[utf8]{inputenc}
\usepackage{graphics}  
\usepackage{xcolor} 
\usepackage{amsmath}  
\usepackage{amsthm}  
\usepackage{amsfonts}  
\usepackage{amssymb} 
\usepackage{bm} 

\usepackage{slashed}
\usepackage{braket}

\usepackage{hyperref}
\usepackage{longtable}
\usepackage{caption}
\usepackage{subfig}

\usepackage{tikz}

\usepackage{cases}

\usepackage{float}

\numberwithin{equation}{section}

\subheader{Addendum}
\title{\boldmath $|V_{cb}|$ and $\gamma$ from $B$-mixing -- Addendum to
``$B_s$ mixing observables and $|V_{td}/V_{ts}|$ from sum rules``}

\author[a]{Daniel~King,}
\author[b]{Matthew~Kirk,}
\author[a]{Alexander~Lenz}
\author[c]{and Thomas~Rauh}

\affiliation[a]{IPPP, Department of Physics,  University of Durham,  \\
  DH1 3LE, United Kingdom}
\affiliation[b]{Dipartimento di Fisica, Università
  di Roma “La Sapienza” \& INFN Sezione di Roma,  \\
  Piazzale Aldo Moro 2, 00185 Roma, Italy
}
\affiliation[c]{Albert Einstein Center for Fundamental Physics, \\
  Institute for Theoretical Physics, University of Bern, \\
  Sidlerstrasse 5, CH-3012 Bern, Switzerland
}

\emailAdd{daniel.j.king@durham.ac.uk}
\emailAdd{matthew.kirk@roma1.infn.it}
\emailAdd{alexander.lenz@durham.ac.uk}
\emailAdd{rauh@itp.unibe.ch}

\abstract{In this addendum to ``$B_s$ mixing observables and $|V_{td}/V_{ts}|$ from sum rules'' \cite{King:2019lal}
we study the impact of the recent improvements in the theoretical precision of $B$ meson mixing 
onto CKM unitarity fits. Our key results are the most precise determination of the angle 
$\gamma = \left(63.4\pm0.9\right)^\circ$ in the unitarity triangle and a new value for the CKM 
element $|V_{cb}|=(41.6\pm0.7)\cdot10^{-3}$.}


\notoc 

\begin{document}
\allowdisplaybreaks
\maketitle
\null\hfill IPPP/19/83
\newpage

\noindent
In our recent works we have determined the hadronic matrix elements for $B$-mixing with HQET sum 
rules~\cite{Kirk:2017juj,King:2019lal} (cf. also~\cite{Grozin:2016uqy}) and combined the results with 
lattice determinations~\cite{Bazavov:2016nty,Boyle:2018knm,Dowdall:2019bea} to obtain updated 
predictions~\cite{DiLuzio:2019jyq} for the mass differences $\Delta M_d$ and $\Delta M_s$. 
Here we use the weighted averages for the matrix elements presented in~\cite{DiLuzio:2019jyq} to 
determine the following combinations of CKM elements 
\begin{eqnarray}
  \label{eq:Vts}  
    |V_{ts} V_{tb}| & = & \left(  40.91^{+0.67}_{-0.64} \right) \cdot 10^{-3}
    \\
    & = &
    \left( 40.91 
      \left.^{+0.65}_{-0.62}\right|_{f_B^2 B}
      \left. \pm 0.17       \right|_{m_t} 
      \left. \pm 0.05\right|_{\alpha_s(M_Z)}
      \left. \pm 0.02 \right|_{\Delta M_s}\right) \cdot 10^{-3}\, ,
    \nonumber
    \\
  \label{eq:VtdoverVts}
  \left| \frac{V_{td}}{V_{ts}} \right| & = & 0.2043^{+0.0010}_{-0.0011}
  \\
  & = & 
   0.2043 
   \left.^{+0.0009}_{-0.0010} \right|_{\xi} 
   \left. \pm 0.0003 \right|_{\Delta M_d} 
   \left. \pm 0.0001 \right|_{\Delta M_s}\, ,
  \nonumber
  \end{eqnarray}
from the experimental measurements of the mass differences, updating the results in~\cite{King:2019lal}. 
As discussed in~\cite{King:2019lal,DiLuzio:2019jyq}, the small theory uncertainty on $|V_{td}/V_{ts}|$ is 
due to the combination of recent lattice results \cite{Aoki:2019cca,Hughes:2017spc,Bazavov:2017lyh}
for the ratio $f_{B_s}/f_{B_d}$ and the precise sum rule results~\cite{King:2019lal} for the ratio of 
the Bag parameters which yields the most precise result for the ratio $\xi$~\cite{DiLuzio:2019jyq}. 
Motivated by the well-known discrepancy between the direct determination of the CKM elements $V_{cb}$ 
and $V_{ub}$ from semi-leptonic $b$-hadron decays (see \cite{Gambino:2019sif} for some recent discussion) 
and the prospect of a measurement of the CKM angle $\gamma$ with an uncertainty of $1.5^\circ$ by 2023 
from the LHCb collaboration~\cite{Bediaga:2018lhg} we study the impact of these values on CKM unitarity 
fits. 

\begin{figure}[b]
    \centering
    \includegraphics[width=\textwidth]{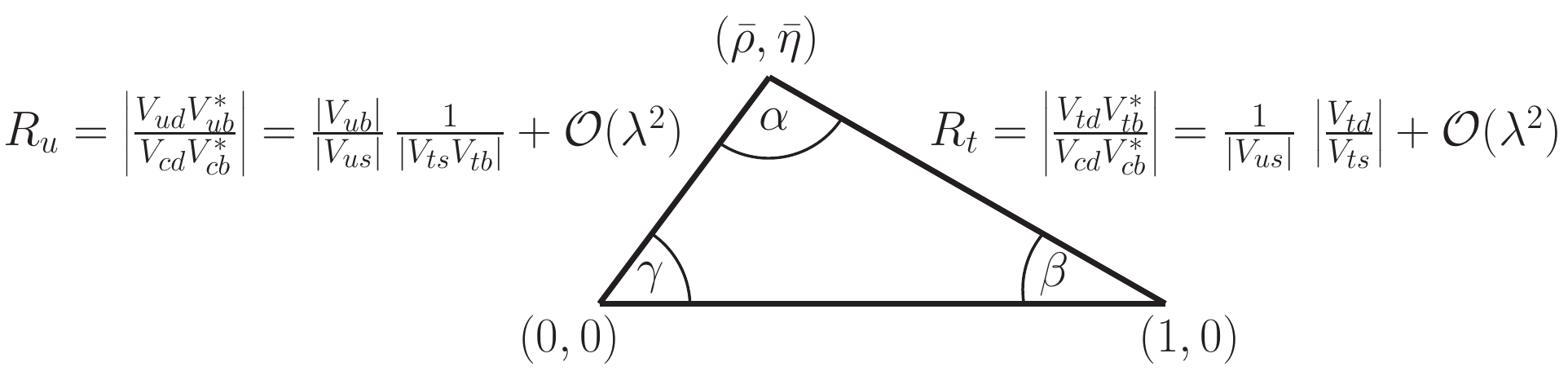}
    \caption{Our conventions for the unitarity triangle in the $\bar{\rho}-\bar{\eta}$ plane.}
    \label{fig:UTlabels}
\end{figure}

The effects of $B$-mixing on CKM unitarity fits can be illustrated with the unitarity triangle 
shown in Figure~\ref{fig:UTlabels}. The combinations of CKM elements \eqref{eq:Vts} and \eqref{eq:VtdoverVts} 
we determined from $\Delta M_s$ and $\Delta M_d$ appear in the lengths of the two non-trivial 
sides of the triangle if we expand to leading order in the Wolfenstein parameter $\lambda=|V_{us}|$. 
Up to reflection with respect to the $\bar{\rho}$ axis the apex of the triangle is exactly fixed 
with the addition of $|V_{ub}|$ and the precisely measured $|V_{us}|$. Here, we use this information 
to determine the angle $\gamma$. Furthermore, we can extract $|V_{cb}| = |V_{ts}V_{tb}|\times[1+\mathcal{O}(\lambda^2)]$ 
with a precision that is competitive with direct measurements.  

\begin{figure}[H]
    \centering
    \includegraphics[width=0.95\textwidth]{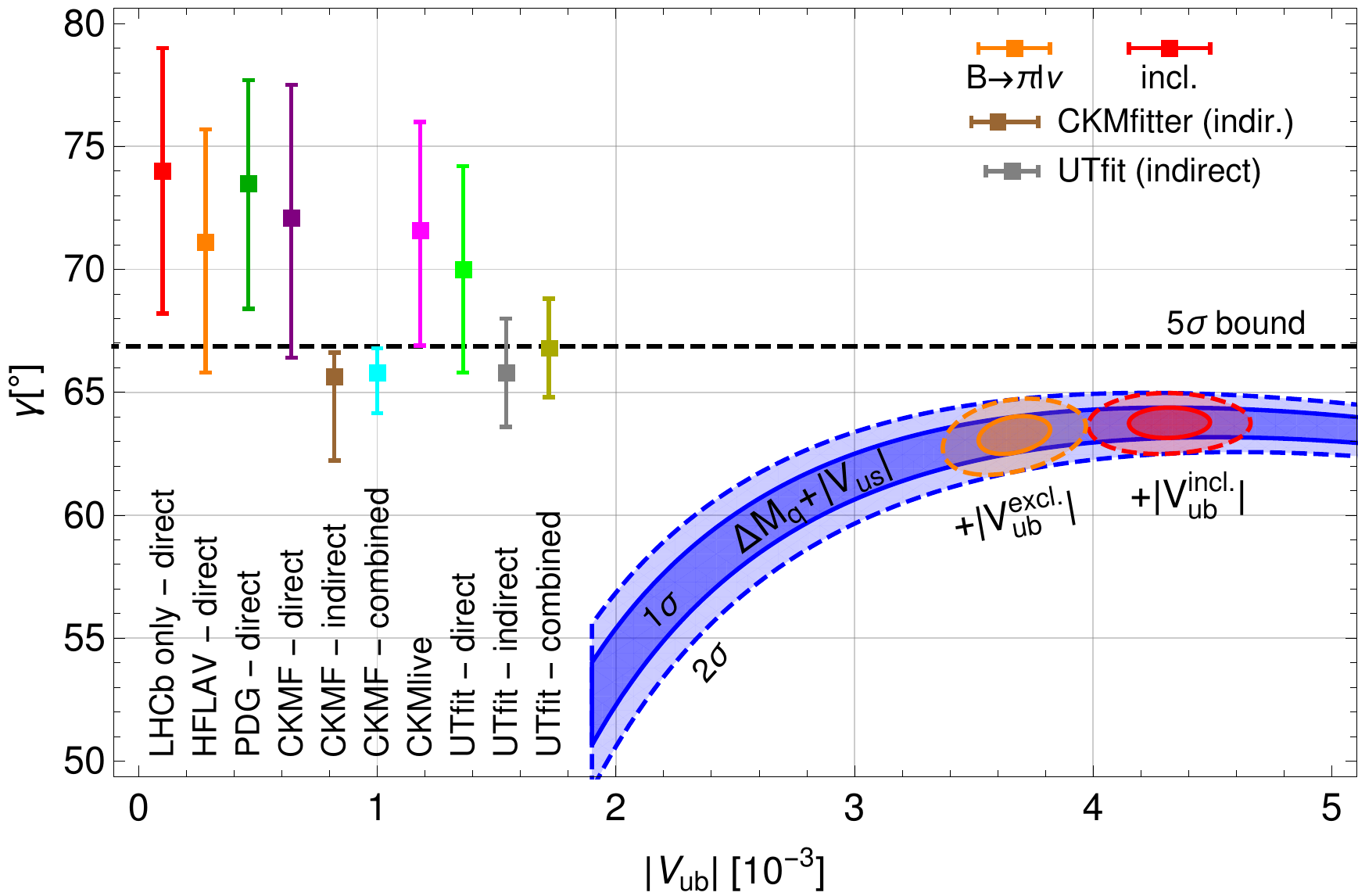}\\[0.5cm]
    \includegraphics[width=0.95\textwidth]{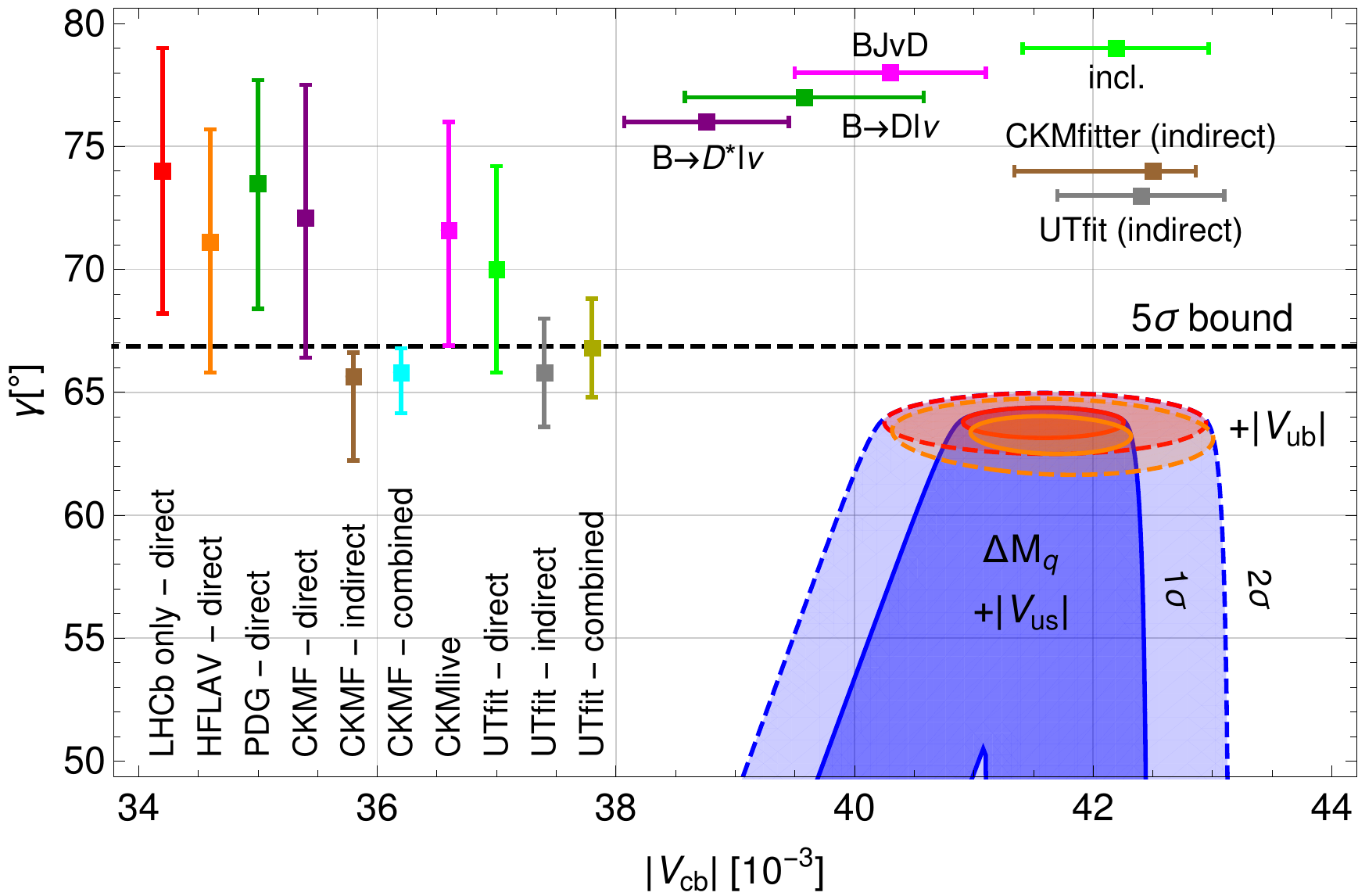}
    \caption{Our results for a minimalistic CKM unitarity fit based on direct measurements of 
    $|V_{us}|$ and the mass differences $\Delta M_d$ and $\Delta M_s$ are given as shaded blue
    regions. Including the exclusive or inclusive measurements of $|V_{ub}|$ yields the orange 
    and red regions, respectively. See text for details.}
    \label{fig:fits}
\end{figure}

We perform a minimalistic CKM unitarity fit (cf. the appendix for a description of the fit procedure), 
first taking only the direct measurements of the CKM element $|V_{us}| = 0.2243\pm0.0005$ 
\cite{Tanabashi:2018oca} and the mass differences $\Delta M_d$ and $\Delta M_s$ into account. 
This strongly constrains the length of the side $R_t$. Figure~\ref{fig:fits} shows our results in 
the $|V_{ub}|-\gamma$ and $|V_{cb}|-\gamma$
\begin{figure}
    \centering
    \includegraphics[width=0.6\textwidth]{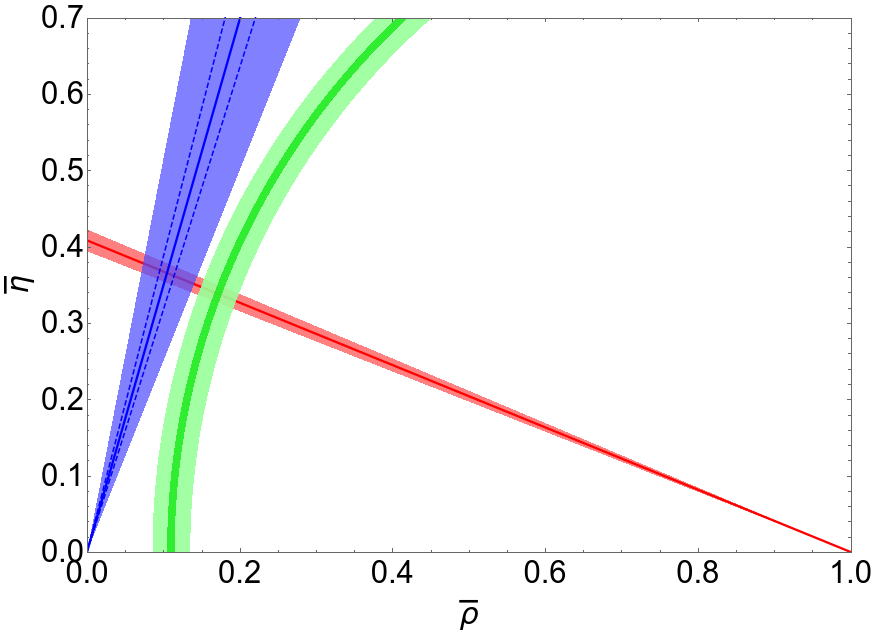}
    \caption{We show the constraints on the apex of the unitarity triangle from the direct 
    measurement of $\gamma$ from LHCb~\cite{Kenzie:2018oob} (blue), $B$-mixing (green) and 
    the value of $\beta$, taken from HFLAV~\cite{Amhis:2019ckw} (red). The dark and light 
    green regions indicate the $1\sigma$ and $5\sigma$ bounds, while the blue and red regions 
    refer to the $1\sigma$ constraints. The dashed blue lines illustrate the future precision 
    of $\pm1.5^\circ$ on the measurement of $\gamma$. See \cite{Blanke:2018cya} for a version 
    of this plot prior to the recent improvements in the theory predictions for the mass 
    differences.} 
    \label{fig:Gamma}
\end{figure}
\noindent planes where the shaded blue regions indicate the 
parameter space satisfying the inputs within one and two standard deviations. Obviously, the CKM fit 
with only three inputs is underconstrained which is reflected by the fact that the blue region traces 
a one-dimensional best-fit path in the 4-dimensional parameter space of the CKM matrix. 
Nevertheless, the underconstrained fit is sufficient to obtain an important constraint. Namely, 
for values of $\gamma$ larger than about $65^\circ$ the unitarity triangle does not close within 
the two-sigma region -- independently of the value of the unconstrained degree of freedom which 
corresponds to the length of the side $R_u$ in the unitarity triangle\footnote{
Similar observations were made in e.g.\cite{Blanke:2016bhf,Blanke:2018cya}.}. 
This behaviour is illustrated in Figure~\ref{fig:Gamma} and allows us to derive a stringent upper 
limit on $\gamma$. At the level of five standard deviations we obtain 
\begin{equation}
 \gamma\leq66.9^\circ\hspace{1.5cm}[5\sigma]\,, 
 \label{eq:gamma_bound}
\end{equation}
which is indicated by the horizontal dashed line in Figure~\ref{fig:fits} and quite a bit smaller 
than the direct measurements of $\gamma$~\cite{Kenzie:2018oob,Amhis:2019ckw,Tanabashi:2018oca,Charles:2004jd,Bona:2006ah} 
summarised there. We note that the indirect determinations of $\gamma$ from the 
CKMfitter~\cite{Charles:2004jd} and UTfit~\cite{Bona:2006ah} collaborations also yield smaller values 
than direct measurements, albeit larger ones than our analysis. We used the CKMlive \cite{CKMlive} tool 
to perform the standard CKMfitter analysis without direct measurements of $\gamma$ or the mass differences 
and obtained the result 
\begin{equation}
 \gamma = \left( 71.6^{+4.4}_{-4.7} \right)^\circ, \hspace{1.5cm} \text{CKMlive -- fit without $\gamma, \Delta M_s, \Delta M_d$} \,, \label{gamma:CKMlive}
\end{equation}
which is in good agreement with the direct measurements of $\gamma$ and has a significantly larger 
uncertainty than the indirect fit results. This demonstrates that the smaller indirect values in the 
CKMfitter and UTfit studies are solely driven by $\Delta M_s$ and $\Delta M_d$ and implies that 
the confrontation of the planned improvements by LHCb and Belle II for the experimental determination 
of $\gamma$ with constraints from the mass differences is a very promising indicator for BSM 
physics. Assuming the central value of the direct measurement remains the expected precision of 
$\pm1.5^\circ$ by 2023 will lead to a significant tension as indicated in Figure~\ref{fig:Gamma}. 

For smaller values of $\gamma$ there are two intersections between the circle of length $R_t$ 
around the point (1,0) and the line crossing the origin at angle $\gamma$, leading to two 
degenerate perfect-fit results for $|V_{ub}|$ and $|V_{cb}|$ at a fixed value of $\gamma$. 
This degeneracy can be broken by constraining the length of the side $R_u$ by including the 
measurements of $|V_{ub}|$ in the fit. Due to the well-known puzzle about different results 
in exclusive and inclusive measurements (shown by the orange and red horizontal error bars 
in Figure~\ref{fig:fits}, values from HFLAV~\cite{Amhis:2019ckw}) this step would normally 
have to be taken with a grain of salt. However, due to a lucky numerical coincidence the 
values of $|V_{ub}|$ are very close to the region where the intersection point of the circles 
of length $R_t$ and $R_u$ lies at the maximal value of $\gamma$ allowed by $R_t$ as shown by 
the orange and red ellipses in  Figure~\ref{fig:fits} which are the results of the fit when 
the exclusive or inclusive measurements of $|V_{ub}|$ are included. Thus, the dependence of 
$\gamma$ on the exact value of $|V_{ub}|$ is rather small. Indeed we find 
\begin{align}
 \gamma ={}& \left(63.3_{-0.8}^{+0.7}\right)^\circ\,,\hspace{1cm}\text{from }|V_{ub}^\text{excl.}|\,,\\
 \gamma ={}& \left(63.8_{-0.6}^{+0.6}\right)^\circ\,,\hspace{1cm}\text{from }|V_{ub}^\text{incl.}|\,. 
\end{align}
We take the envelope of both values 
\begin{equation}
 \gamma = \left(63.4\pm0.9\right)^\circ\,, 
 \label{eq:gamma}
\end{equation}
as our final result to be sufficiently conservative about the uncertainty associated with the 
direct measurements of $|V_{ub}|$. Eq.~\eqref{eq:gamma} represents the most precise determination 
of $\gamma$ to date. The result is fairly insensitive to the input value for $|V_{us}|$. 
If we inflate the error in $|V_{us}|$ by a factor of three we obtain $\gamma = (63.4\pm1.3)^\circ$ 
and the upper five-sigma bound \eqref{eq:gamma_bound} becomes $68.9^\circ$. A similar value of 
$\gamma=(63.4\pm1.5)^\circ$ with an upper five-sigma bound of $69.4^\circ$ results from doubling 
the theory uncertainty on the ratio $\xi$. Even in both of these more conservative scenarios, 
a future LHCb measurement of $\gamma$ with an accuracy of $1.5^\circ$ and an unchanged central 
value would still correspond to a tension at the level of five standard deviations. This demonstrates 
the robustness of refined direct measurements of $\gamma$ as a probe of new physics effects. 

The effect of the exclusive or inclusive $|V_{ub}|$ measurements on the fit is also indicated 
in the $|V_{cb}|-\gamma$ plane  by the orange and red ellipses, respectively. The difference in 
the extracted values of $|V_{cb}|$ is negligible and we again adopt the envelope as our final result 
\begin{equation}
 |V_{cb}| = \left(41.6\pm0.7\right)\cdot10^{-3}\,.
 \label{eq:vcb}
\end{equation}
We also show the exclusive and inclusive HFLAV averages~\cite{Amhis:2019ckw} and the result of 
a recent reanalysis BJvD~\cite{Bordone:2019vic} of exclusive determinations in Figure~\ref{fig:fits}. 
Our result yields a competitive uncertainty and the one-sigma region overlaps with the inclusive 
and the BJvD results, while there is a 1.7 and 2.9 $\sigma$ tension with respect to the $B\to D\ell\nu$ 
and $B\to D^*\ell\nu$ values quoted by HFLAV. The result \eqref{eq:vcb} remains unaffected when we inflate 
the $|V_{us}|$-uncertainty by a factor of three or the theory uncertainty for $\xi$ by a factor of two. 

In summary, we have performed a minimal $\chi^2$ fit of the CKM parameters based on the mass 
differences in the $B$ system and direct measurements of $|V_{us}|$ and $|V_{ub}|$. We found  
competitive results for $|V_{cb}|$ which are in good agreement with the inclusive determinations 
and obtained the currently most precise value for the angle $\gamma$ in the unitarity triangle. 
Our analysis clearly shows that more precise measurements of $\gamma$ are a sensitive probe of 
new physics effects in the flavour sector. We are looking forward to updates of the complete CKM 
unitarity fits by the CKMfitter and UTfit collaborations where the latest theoretical developments 
\cite{Grozin:2016uqy,Kirk:2017juj,King:2019lal,Bazavov:2016nty,Boyle:2018knm,Dowdall:2019bea,DiLuzio:2019jyq} 
in $B$-mixing are taken into account.

\acknowledgments{The work of D.K. and A.L.  was supported by the STFC through the IPPP grant and a
Postgraduate Studentship.
M.K.\ was supported by MIUR (Italy) under a contract PRIN 2015P5SBHT and by INFN Sezione di Roma La Sapienza and partially supported by the ERC-2010 DaMESyFla Grant Agreement Number: 267985.
}

\vspace{0.5cm}
\noindent\textbf{Note added:} The latest LHCb measurement of $|V_{cb}|$ from exclusive semileptonic 
$B_s$ decays~\cite{Aaij:2020hsi} is in excellent agreement with our indirect result \eqref{eq:vcb}. 

\vspace{0.5cm}
\noindent\textbf{Appendix: Description of the fit procedure}\\[0.5cm]
We perform the fit in the standard parametrization of the CKM triangle with the three angles 
$\theta_{12}$, $\theta_{13}$ and $\theta_{23}$ as well as the phase $\delta$ which will be 
denoted by the four-dimensional vector $\bm{\theta}$ below. No approximations related to the 
Wolfenstein parametrization are made in the numerical analysis. We define a $\chi^2$ function as 
\begin{equation}
 \chi^2(\bm{\theta}) = \sum_{x\in X}\left(\frac{x(\bm{\theta}) - x_\text{in}}{\Delta x}\right)^2
\end{equation}
where the underconstrained fit, corresponding to the blue regions in Figure~\ref{fig:fits}, utilizes 
$X=\{|V_{ts}V_{tb}|,|V_{td}/V_{ts}|,|V_{us}|\}$ and the orange and red regions are obtained by 
including $|V_{ub}|$ in $X$. The $x_\text{in}$ and $\Delta x$ correspond to the central values 
and total uncertainties of these quantities which are given in Eq.~\eqref{eq:Vts}, 
Eq.~\eqref{eq:VtdoverVts} and the text. The contours in Figure~\ref{fig:fits} are then obtained 
by scanning over the parameter space, e.g. the one-sigma regions in the $|V_{cb}|-\gamma$ plane 
corresponds to all points $\vec{x}$ which satisfy 
\begin{equation}
 \mathop{\text{Min}}\limits_{\bm{\theta}\text{ s.t. }(|V_{cb}|(\bm{\theta}),\gamma(\bm{\theta})) \,=\, \vec{x}}\left(\chi^2(\bm{\theta})\right) \leq 1 \, .
\end{equation}

\bibliography{AddendumCKM}
\bibliographystyle{JHEP}

\end{document}